\documentclass[10pt,final,conference,letterpaper,romanappendices,twocolumn]{IEEEtran}
\usepackage{graphics}
\usepackage{cite}
\usepackage{epsfig}
\usepackage{enumerate}
\usepackage{amsthm}
\usepackage{amsmath,amssymb,amsfonts,amstext,amsbsy,amsopn,dsfont}
\usepackage{cases}
\usepackage{sublabel}
\usepackage{array}

\newtheorem{lemma}{\textbf{Lemma}}

\newcommand{\defn}{\triangleq}
\newcommand{\dif}{\textmd{d}}

\begin{document}

\title{ Optimal Base Station Deployment for Small Cell Networks with Energy-Efficient Power Control}
\author{Ching-Ting Peng, Li-Chun Wang, Chun-Hung Liu \\Department of Electrical and Computer Engineering \\National Chiao Tung University\\ Hsinchu, Taiwan}
\maketitle

\begin{abstract}
In this paper, how to optimally deploy base station density in a small cell network with energy-efficient power control was investigated. Base stations (BSs) and users form two independent Poisson point processes (PPPs) in the network. Since user-centric cell association may lead to void cells that do not have any users, the power of each BS is controlled in either all-on or on-off mode depending on whether its cell is void or not. The average cell rates for each power control mode are first found and their corresponding energy efficiency is also characterized.  The optimal BS density that maximizes the energy efficiency under a given user density is theoretically proved to exist and its value can be found numerically. Both analytical and simulated results indicate that on-off power control is significantly superior to all-on power control in terms of energy efficiency if BSs are deployed based on their optimal energy-efficient density. 
\end{abstract}

\section{Introduction}
In recent years, we have witnessed a trend that mobile devices, such as smart phones and tablets, have been proliferating and relentlessly penetrating our daily life.  Such powerful handset devices have created a new dimension of transmitting information over newly developed wireless technologies. Accordingly, this situation has urged customers to expect more and more data rate to satisfy their versatile wireless demands. To boost network throughput, one of the most effective approaches is to increase the cell density such that communication links can be created as many as possible and the area spectral efficiency (bps/Hz/km${}^2$) is accordingly improved. Therefore a small cell network consisting of picocells or femtocells has gained much attention since such small cells can be deployed indefinitely and the throughput gain they provide seems not to have a hard limit.  

\subsection{Motivation and Prior Work}
Although small cells can increase the overall network capacity, there arise some problems regarding them, such as inter-cell interference mitigation and power control. How many of BSs should be deployed and working in a given area is a pivotal network design issue since an inappropriate BS deployment could give rise to unpleasant outcomes, such as excess interferences, no users in cells (called void cells), high power consumption and system operating cost, etc.. For user-centric cell association, the void cell issue in a small cell network is especially serious since the BS density is not significantly small compared to the user density \cite{CHLLCW15}, and how to optimally deploy BSs by considering the void cell issue seems not to have a good guideline to follow until now. This motivates us to study if there exists an optimal BS deployment density for a small cell network with void cells. 

In this paper, all BSs in a small cell network are assumed to form a homogeneous Poisson point process (PPP) and they use \textit{all-on} or \textit{on-off} power control schemes -- a BS with the all-on power control is always turned on even in the situation that it does not have any users to serve, whereas a BS with the on-off power control is turned on only when it has to serve at least one user, otherwise it is off. The on-off power control is suitable for a small cell network since  small cells typically consume much less power and can be dynamically switched on or off. Thus, a small cell network can attain the energy-efficient goal much easier than a traditional cellular network\cite{hwa13,ash11}.

Previous theoretical works on the optimal deployment density of BSs in a heterogeneous or small cell network are still fairly minimal. Reference \cite{que11} only numerically showed that an optimal pico-macro density ratio that maximizes the defined energy efficiency exists. In \cite{cho13}, with the constraints on the average user rate, the authors studied the minimum distance between small cells that minimizes the power consumption of a two-tier heterogeneous network. In \cite{cao13}, a closed-form upper bound on the optimal cell density that satisfies some constraints on the user outage rate is found but the energy efficiency of BSs is not considered in this work. References \cite{que11,cho13,cao13} did not consider that BSs can lower transmit power in a densely deployed small cell network (an interference-limited network) to improve energy efficiency since the coverage probability and rate are not sensitive to the changes of the BS transmit power. Also, these prior works do not study the optimal BS deployment density from an achievable rate point of view and they overlook the void cell issue in their models. 

\subsection{Contributions}

Since there are very few works of Poisson cellular networks that notice the void cell issue in the model\footnote{The recent work in \cite{CHLLCW15} has a more complete study on the cell voidness problem in a dense small cell network.}, our first contribution in this paper is to characterize the void probability of a cell and model the interference with the void cell phenomenon, and propose all-on and on-off power control schemes for a BS. The second contribution is to find the average achievable user and cell rates in the network under the void cell issue and use them to characterize the energy efficiency of a BS for all-on and on-off power controls by using a proper power consumption model of a BS in the literature. Although the average maximum achievable user rate is already derived in \cite{and11} for a given BS density, it does not include the impact of the void cells. In \cite{cho13,jo12}, the average user rate is defined as the average maximum achievable rate divided by the average number of users in a cell such that it scales linearly with the BS density, which makes the analytic results in \cite{cho13,jo12} be detached from the reality since no void cells are considered in the model as well. An optimization problem  of the BS density for the proposed power control is formulated based on the BS power efficiency, and it is shown that the optimal BS density that maximizes the power efficiency certainly exists for a given user density and can be found numerically, which is our third contribution. Numerical results verified that the power efficiency achieved by on-off power control significantly outperforms that achieved by all-on power control if the found optimal BS density is adopted.   

\section{System Model}
\begin{figure}[!t]
\centering
\includegraphics[width=3.5in, height=2.65in]{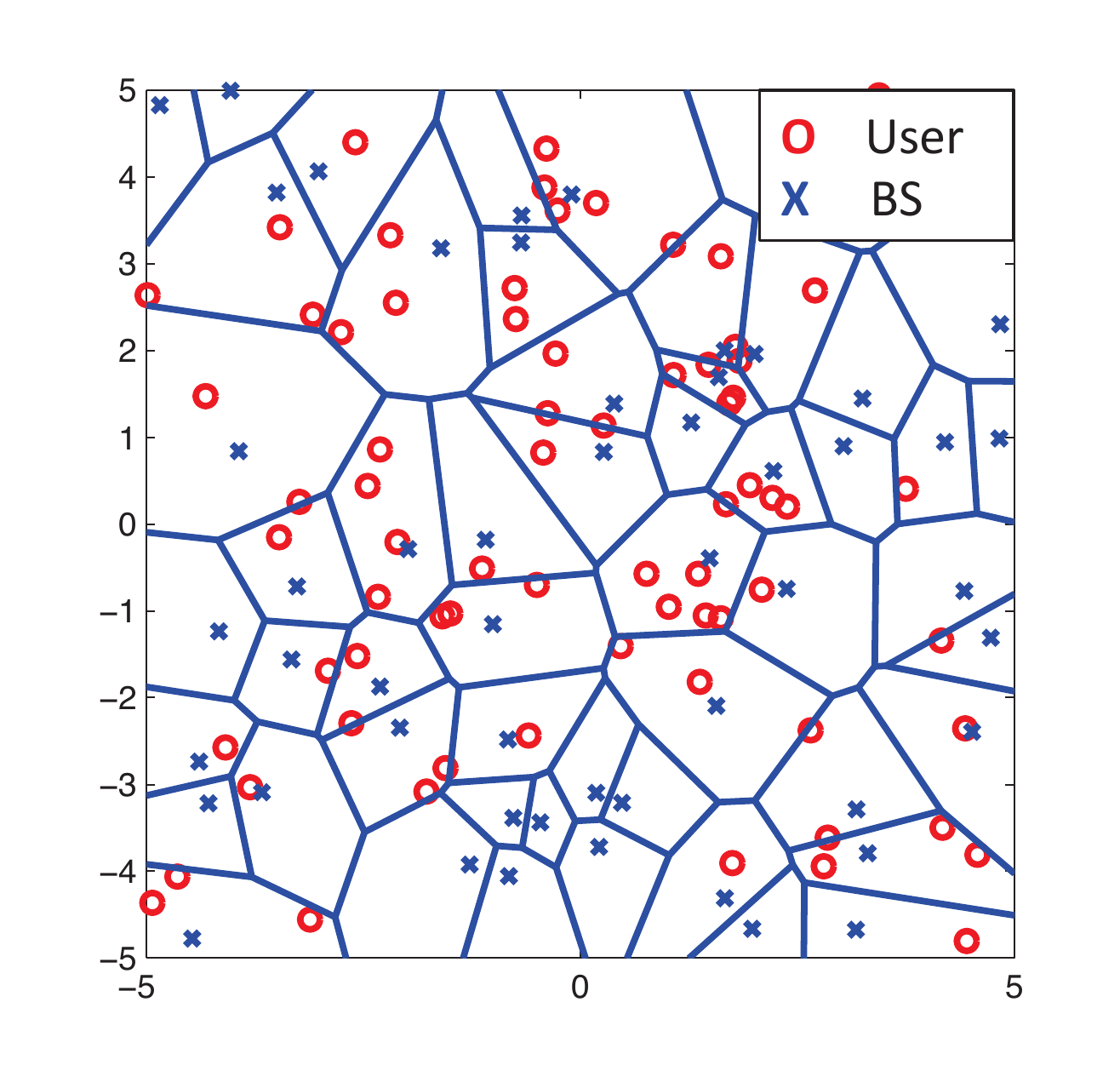}
\caption{A simulation example of the distributions of BSs and users in a single-tier small cell network}
\label{fig:voronoi_cell_with_ue}
\end{figure}

In a small cell network, we consider all BSs form a homogeneous PPP $\Phi_b$ of density $\lambda_b$, and all users in the network form another independent homogeneous PPP $\Phi_u$ of density $\lambda_u$ (see an illustration in Fig. \ref{fig:voronoi_cell_with_ue}). We assume there is no inter-cell cooperation and the interferences from other macro-cells are negligible. The powers of transmitted signals undergo path loss and Rayleigh fading. Thus if the distance between a user and a BS is $R$ then the channel power gain model can be expressed as $HCR^{-\alpha}$ where $H$ is due to Rayleigh fading and it is an exponential random variable with unit mean and variance, $C>0$ is a channel-dependent constant and $\alpha>2$ is the path loss exponent.

Let $P_t$ denote the transmit power of a BS and it should be properly kept in an appropriately  low level since in an interference-limited network, such as a cellular network, the maximum achievable rate is not sensitive to the changes of power levels and thus using high transmit powers in this context is not energy-efficient. Usually, $P_t$ is determined by referring to the BS density since a BS in a densely deployed network (e.g. picocells or fentocells) typically has a lower transmit power  than a BS in a sparsely deployed network (e.g. macro cells). Accordingly, in a dense small cell network, reducing transmit powers of BSs should attain a  higher power efficiency.  Specifically, we heuristically select $P_t$ that makes the probability that (long-term) received signal powers are smaller than a predesignated minimum power $P_{r,\min}$ be less than a constant $\delta\in(0,1)$. In other words, $P_t$ has to satisfy the following inequality constraint: 
\begin{equation}\label{Eqn:TxPowCon}
\mathbb{P}\left[HCP_tR^{-\alpha}\leq P_{r,\min}\right]\leq \delta.
\end{equation}
The following lemma gives the lower bound on $P_t$ if a user is always associated with its nearest BS.
\begin{lemma}
If all users associate with their nearest BS, the minimum transmit power $P_t$ satisfying \eqref{Eqn:TxPowCon} is given by
\begin{equation}\label{Eqn:LowBoundTxPow}
P_t = \frac{P_{r,0}}{C\lambda_b^{\frac{\alpha}{2}}},
\end{equation}
where $P_{r,0}=(-\ln\delta/\pi\Gamma(1+\frac{2}{\alpha}))^{\frac{\alpha}{2}}P_{r,\min}$,  and $\Gamma(x)=\int_{0}^{\infty}t^{x-1}e^{-t}\dif t$ is the Gamma function.
\begin{IEEEproof}
The expression in \eqref{Eqn:TxPowCon} for a given $H$ can be rewritten as
$$\mathbb{P}\left[R>\left(\frac{HCP_t}{P_{r,\min}}\right)^{\frac{1}{\alpha}}\bigg| H\right]=e^{-\pi\lambda_b \left(\frac{HCP_t}{P_{r,\min}}\right)^{\frac{2}{\alpha}}}\leq \delta,\,\forall H>0,$$
where the equality follows from the fact that the BS always connects to its nearest BS and the complementary cumulative density function (CCDF) of $R$ is given in \cite{FBBBL10}. Thus $P_t$ can be expressed as
$$P_t \geq \frac{P_{r,\min}}{C} \left(\frac{-\ln\delta}{\pi \lambda_b\mathbb{E}[H^{\frac{2}{\alpha}}] }\right)^{\frac{\alpha}{2}}\geq \frac{P_{r,0}}{C},$$
where the second inequality follows from Jensen's inequality and $\mathbb{E}[H^{\frac{2}{\alpha}}]=\Gamma(1+\frac{2}{\alpha})$. This completes the proof. 
\end{IEEEproof} 
\end{lemma}

In order to gain some insight on whether transmit powers are effectively utilized in a small cell network or not, in this paper we consider two power control schemes performed by BSs - one is the ``all-on'' scheme and the other is the ``on-off'' scheme. The ``all-on'' scheme means no matter if a BS has a user to serve, it is always turned on, whereas BSs with  the ``on-off'' scheme will be turned off  if they do not have any users and otherwise they remain on. If a BS is off,  it is assumed to at least consume power $P_{\text{off}}$ for functioning itself hardware. On the contrary, if the BS is on, the power it consumes  can be described by the following model proposed in \cite{aue11} 
\begin{equation}
P_{\text{on}}=P_0+\Delta\, P_t,
\end{equation}
where $P_0$ and $\Delta$ are constants depending on BS types. Note that we assume $P_0>P_{\text{off}}$, which is usually the case in practice.

Since there are two power control schemes for BSs, we have to specify the two densities of the BSs for the two power control schemes respectively before characterizing the signal-to-interference plus noise ratio (SINR) of the user. First consider the case that all BSs use the all-on power scheme.  Let $\lambda_1$ be the density of the BSs that are on and in this case $\lambda_1=\lambda_b$. For the on-off power control, since each BS in the network independently decides to turn on or off based on whether it has a user or not, the turned-on BSs is a thinned PPP of the original $\Phi_b$ and their density is given by
\begin{equation}\label{Eqn:BsDensityOnOffPow}
\lambda_2 =\left(1-\mathbb{P}[N_b=0]\right)\lambda_b,
\end{equation}  
where $N_b$ is the number of users served in a cell and thus $\mathbb{P}[N_b=0]$ is the probability that a BS does not have a user.  The distribution of $N_b$ is experimentally characterized in \cite{yu13}, which is
\begin{equation}\label{nb_pmf}
\mathbb{P}[N_b=n]=\frac{3.5^{3.5}\Gamma(n+3.5)\mu^n}{\Gamma(3.5)n!(\mu+3.5)^{n+3.5}},
\end{equation}
where $\mu\defn \lambda_u/\lambda_b$ is called \textit{cell load}, i.e. the average number of users per cell. Substituting \eqref{nb_pmf} with $n=0$ into \eqref{Eqn:BsDensityOnOffPow}, we have
\begin{align}\label{Eqn:Density2}
\lambda_2=\left[1-p_0(\mu)\right]\lambda_b,
\end{align}
where $p_0(\mu)\defn \left(1+\mu/3.5\right)^{-3.5}$. Therefore, the SINR of a user located at the origin can be represented by  
\begin{equation}
\mathtt{SINR}_k=\frac{HC P_tR^{-\alpha}}{\sigma_n^2+\sum_{X_j\in\Phi_k\setminus\{X: \|X\|\leq R\}}CP_tH_j\|X_j\|^{-\alpha}},
\end{equation}
where $H$, $H_j$ are i.i.d. exponentially distributed with unit mean and variance, $\sigma_n^2$ is the noise power, $\|X_j\|$ denotes the Euclidean distance of the BS $X_j$ away from the user, and $\Phi_k$ corresponds to the PPP with density $\lambda_k$, $k\in\{1,2\}$.

\section{Analysis of Cell Rate and User Rate}
In this section, we would like to investigate the power efficiency of a BS in a small cell network. The average maximum achievable rate of a user in the network with a specific power control scheme should be first analyzed since we need it to evaluate the average cell rate that is used to define the power efficiency of a BS. 

If the channel input distribution is Gaussian and all interferences are treated as noise, the average maximum achievable rate (i.e. the Shannon capacity per bandwidth) of a user is $\mathcal{C}_k\defn\mathbb{E}[\log_2(1+\mathtt{SINR}_k)]$. The following lemma gives us the explicit result of $\mathcal{C}_k$ for each power control scheme.
\begin{lemma}
Let ${}_2F_1(a,b;c;z)$ stand for the Gaussian hypergeometric function \cite{abramowitz2012handbook} and use it to define $\rho(T,\alpha)$ as follows
\begin{equation}
\rho(T,\alpha)=\left(\frac{2T}{\alpha-2}\right){}_2F_1\left(1,1-\frac{2}{\alpha};2-\frac{2}{\alpha};-T\right).
\end{equation}
The average maximum achievable rate $\mathcal{C}_1$ of a user for the all-on power scheme is given by
\begin{equation}\label{Eqn:C1}
\mathcal{C}_1 =\pi\int\limits_{0}^{\infty}\int\limits_{0}^{\infty}e^{-\frac{\sigma_n^2(2^t-1)}{P_{r,0}}x^{\alpha/2}-\pi x\left(\rho\left(2^t-1,\alpha\right)+1\right)}\dif x\dif t,
\end{equation} 
and $\mathcal{C}_2$, the average maximum achievable rate of a user for on-off power control, can be shown as
\begin{align}\label{Eqn:C2}
\mathcal{C}_2 =\pi\int\limits_{0}^{\infty}\int\limits_{0}^{\infty}e^{-\frac{\sigma_n^2(2^t-1)}{P_{r,0}}x^{\alpha/2}-\pi x\left([1-p_0(\mu)]\rho\left(2^t-1,\alpha\right)+1\right)}\dif x \dif t.
\end{align} 
\end{lemma}  
\begin{IEEEproof}
 $\mathcal{C}_k$ can be expressed as shown in the following:
\begin{align}
\mathcal{C}_k&=\int\limits_{0}^{\infty}\mathbb{P}[\mathtt{SINR}\geq 2^t-1] \dif t\nonumber\\
&\stackrel{(a)}{=}2\pi\lambda_b\int\limits_{0}^{\infty}\int\limits_{0}^{\infty}r e^{-\frac{\sigma_n^2(2^t-1)}{P_t}r^{\alpha}-\pi r^2\left(\lambda_k\rho\left(2^t-1,\alpha\right)+\lambda_b\right)}\dif r\dif t \nonumber\\
&\stackrel{(b)}{=}\pi\int\limits_{0}^{\infty}\int\limits_{0}^{\infty}e^{-\frac{\sigma_n^2(2^t-1)}{P_{r,0}}x^{\alpha/2}-\pi x\left(\frac{\lambda_k}{\lambda_b}\rho\left(2^t-1,\alpha\right)+1\right)}\dif x\dif t,\label{Eqn:CkIntExp}
\end{align}
where $(a)$ follows from the result in \cite{and11} and $(b)$ is obtained by using the lower bound on $P_t$ in \eqref{Eqn:LowBoundTxPow} and changing variable $\lambda_b r^2$ to $x$. For $k=1$, substituting $\lambda_1=\lambda_b$ into \eqref{Eqn:CkIntExp} and thus $\mathcal{C}_1$ in \eqref{Eqn:C1} is obtained. Similarly, $\mathcal{C}_2$ in \eqref{Eqn:C2} can be shown by plugging $\lambda_k=\lambda_2$ into \eqref{Eqn:CkIntExp}.  
\end{IEEEproof}

For the BSs with all-on power control, $\mathcal{C}_1$ in \eqref{Eqn:C1} does not depend on the BS density and cell load, but on $P_{r,0}$. This indicates that deploying more BSs in a fixed area does not help elevate the average maximum achievable rate of a user in the network. As for the BSs with the on-off power scheme, $\mathcal{C}_2$ in \eqref{Eqn:C2} reveals that it is a decreasing function of $\mu$, and $\mathcal{C}_2$ will increase up to a constant, i.e. as $\mu\approx 0$ we have
\begin{align}
 \mathcal{C}_2\lessapprox\pi\int_{0}^{\infty}\int_{0}^{\infty}e^{-\frac{\sigma^2_n(2^t-1)}{P_{r,0}}x^{\frac{\alpha}{2}}-\pi x}\dif x\,\dif t.
\end{align}
That means the maximum achievable rate of a user is barely improved and affected by deploying more BSs if the user density is already significantly smaller than the BS density.

The outage probability $\mathcal{P}_k\triangleq\mathbb{P}[\mathtt{SINR}_k<T]$ also has a similar situation because it can be calculated as follows
\begin{align}
\mathcal{P}_k=1-\pi\int_{0}^{\infty}\int_{0}^{T}e^{-\frac{\sigma_n^2(2^t-1)}{P_{r,0}}x^{\frac{\alpha}{2}}-\pi x\left(\frac{\lambda_k}{\lambda_b}\rho\left(2^t-1,\alpha\right)+1\right)}\dif x\dif t\label{eq:outage_1},
\end{align}
which also indicates that $\mathcal{P}_1$ does not depend on the BS density and cell load and $\mathcal{P}_2$ depends on the cell load only. Note that $\mathcal{P}_k$ is a monotonically decreasing function of $P_{r,0}$ and thus we are able to choose a proper value of $P_{r,0}$ such that $\mathcal{P}_k$ is kept in an acceptable value of the network regardless of BS density.

To evaluate the total user rate of a cell in a more accurate way, we propose a formula of the average cell rate to characterize the total user rate of a cell. We define the average cell rate for any power control scheme as follows
\begin{equation}\label{Eqn:AvgCellRate}
\mathcal{R}_{\text{cell},k} \triangleq\mathbb{E}\left[\frac{1}{N_b}\sum_{i=1}^{N_b}\log_2(1+\mathtt{SINR}_{k,i})\right],\,k\in\{1,2\},
\end{equation}
where $\mathtt{SINR}_{k,i}$ is the $\mathtt{SINR}_k$ of the $i$th user in a cell. The average cell rate in \eqref{Eqn:AvgCellRate} can be explicitly carried out in the following lemma.
\begin{lemma}
Suppose the average maximum achievable rate $\mathcal{C}_k=\mathbb{E}[\log_2(1+\mathtt{SINR}_k)]$ in a cell is independent of the number of the users in that cell. The average cell rate in \eqref{Eqn:AvgCellRate} can be shown to be
\begin{equation}\label{Eqn:AvgCellRate2}
\mathcal{R}_{\text{cell},k}=\left[1-p_0(\mu)\right]\mathcal{C}_k,\quad k\in\{1,2\},
\end{equation}
where $\mathcal{C}_k$ is given in \eqref{Eqn:CkIntExp}.
\end{lemma} 
\begin{IEEEproof}
Using the law of total expectation, $\mathcal{R}_k$ can be expressed as
\begin{align}
\mathcal{R}_{\text{cell},k}
&=\sum_{n=1}^{\infty}\mathbb{P}[N_b=n]\mathbb{E}\left[\frac{1}{N_b}\sum_{i=1}^{N_b}\log_2(1+\mathtt{SINR}_{k,i})\,\Big|\,N_b=n\right]\nonumber\\
&\stackrel{(c)}{=}\sum_{n=1}^{\infty}\mathbb{P}[N_b=n]\mathbb{E}\left[\log_2(1+\mathtt{SINR}_{k,i})\,|\,N_b=n\right]\nonumber\\
&\stackrel{(d)}{=}\sum_{n=1}^{\infty}\mathbb{P}[N_b=n]\mathbb{E}\left[\log_2(1+\mathtt{SINR}_{k,i})\right]\nonumber\\
&=\left(1-\mathbb{P}[N_b=0]\right)\mathbb{E}\left[\log_2(1+\mathtt{SINR}_{k,i})\right]\nonumber\\
&=\left[1-p_0(\mu)\right]\mathcal{C}_k, \nonumber
\end{align}
where (c) holds since $\mathtt{SINR}_{k,i}$ for all $i$ are identically distributed and (d) follows from the assumption that $\mathtt{SINR}_{k,i}$ and $N_b$ are independent \cite{sin13,cao13}. Then using  $\mathbb{P}[N_b=0]=p_0(\mu)$ and the result in \eqref{Eqn:CkIntExp}, \eqref{Eqn:AvgCellRate2} is obtained. 
\end{IEEEproof}
 It can be observed from \eqref{Eqn:AvgCellRate2} that for all-on power control, $\mathcal{R}_{\text{cell},1}$ is a monotonically increasing function of the cell load $\mu$ since $\mathcal{C}_1$ does not depend on $\mu$, whereas in the next section we will show that $\mathcal{R}_{\text{cell},2}$ is a concave function of $\mu$. 
 
Since the maximum achievable rate $\mathcal{C}_k$ does not reflect the fact that the radio resource in a cell is shared by users  either in time or in frequency. To realistically estimate how much per-user rate with time or frequency sharing could be achieved, the average user rate is defined as
\begin{equation}\label{Eqn:AvgUserRate}
\mathcal{R}_{\text{user},k} \triangleq\mathbb{E}\left[\frac{1}{N_u}\log_2(1+\mathtt{SINR}_k)\right]
\end{equation}
assuming the cell resource is equally shared by $N_u\in\mathbb{N}_{+}$ users. Note that in general these two random variables, $N_u$ and $N_b$, are not identically distributed. Similarly, there exists an approximated result of $\mathcal{R}_{\text{user},k}$ as shown in the following lemma.
\begin{lemma}
According to the result in \eqref{Eqn:AvgCellRate2}, the average user rate \eqref{Eqn:AvgUserRate} can be shown to be
\begin{equation}\label{Eqn:AvgUserRate2}
\mathcal{R}_{\text{user},k} = \frac{[1-p_0(\mu)]}{\mu}\mathcal{C}_k,\quad k\in\{1,2\}.
\end{equation}
\end{lemma}
\begin{IEEEproof}
Using the random incidence technique in \cite{larson1981urban}, $\mathbb{P}[N_u=n]$ can be expressed with $\mathbb{P}[N_b=n]$ as
\begin{equation}
\mathbb{P}[N_u=n]=\frac{n}{\mathbb{E}[N_b]}\mathbb{P}[N_b=n],
\end{equation}
Using the law of total expectation, the average user rate can be expressed as
\begin{align*}
\mathcal{R}_{\text{user},k} &=\sum_{n=1}^{\infty}\mathbb{P}[N_u=n]\mathbb{E}\left[\frac{1}{n}\log_2(1+\mathtt{SINR}_k)\,\Big|\,N_u=n\right]\\
&=\sum_{n=1}^{\infty}\frac{n\mathbb{P}[N_b=n]}{\mathbb{E}[N_b]}\mathbb{E}\left[\frac{1}{n}\log_2(1+\mathtt{SINR}_k)\,\Big|\,N_u=n\right]\\
&=\sum_{n=1}^{\infty}\frac{\mathbb{P}[N_b=n]}{\mathbb{E}[N_b]}\mathbb{E}\left[\log_2(1+\mathtt{SINR}_k)\,\Big|\,N_b=n\right].
\end{align*}
Then the result in \eqref{Eqn:AvgUserRate2} follows from the result in the proof of Lemma 3  and $\mathbb{E}[N_b]=\mu$.
\end{IEEEproof}

\begin{figure}[!t]
\centering
\includegraphics[width=3.6in, height=2.25in]{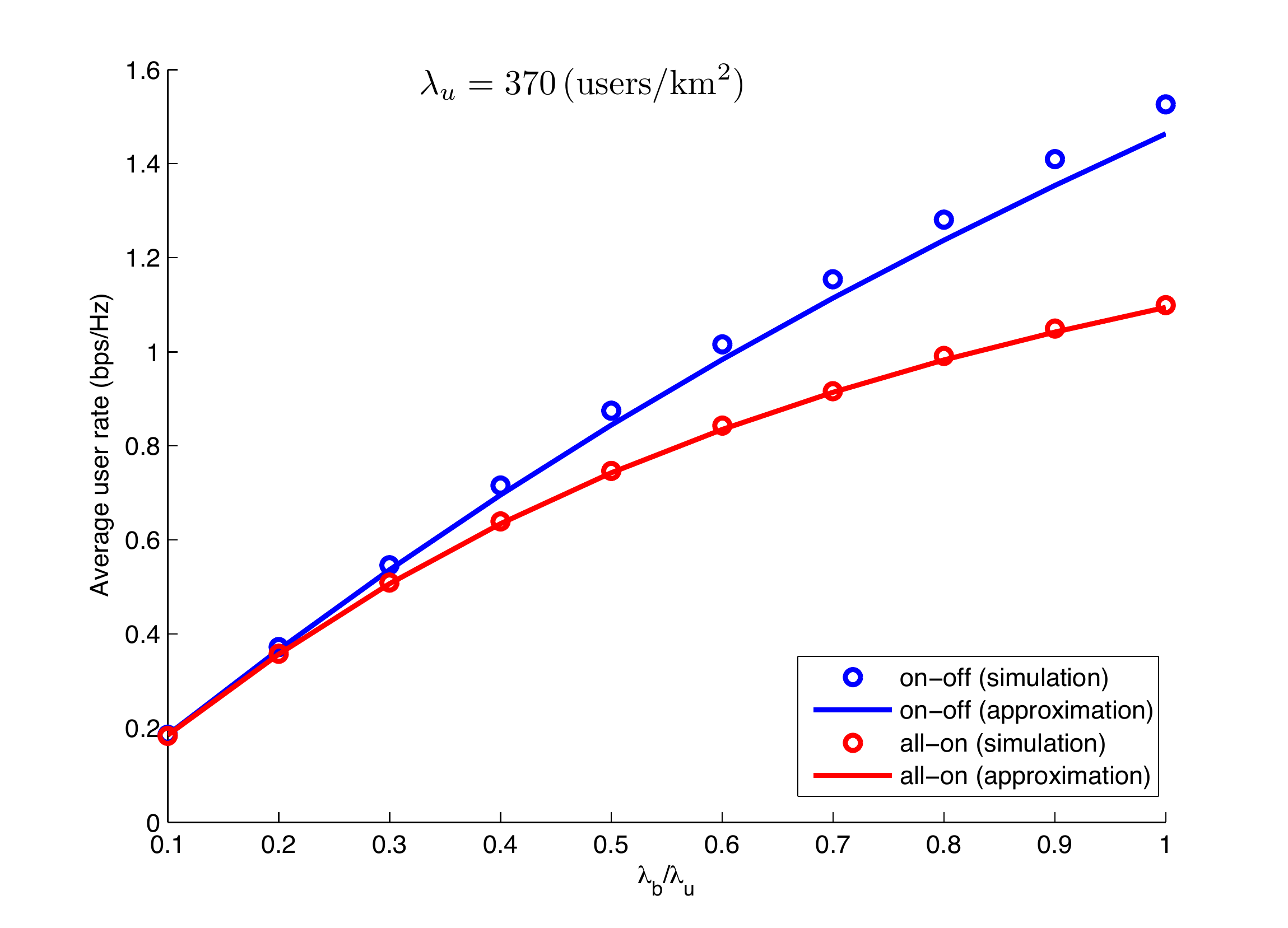}
\caption{Simulation results of verifying the accuracy of the rate result in  \eqref{Eqn:AvgUserRate2}. Network parameters for simulation are the same as the parameters used in Section \ref{Sec:NumericalResults}.}
\label{fig:user_rate}
\end{figure}

As can be observed in \eqref{Eqn:AvgUserRate2}, $\mathcal{R}_{\text{user},k}$ increases as the cell load $\mu$ decreases and it is upper-bounded by a constant
since  both $\mathcal{C}_1$ and $\mathcal{C}_2$ are non-increasing on $\mu$, and  $\frac{1-p_0(\mu)}{\mu}$ is a decreasing function of $\mu$ and goes up to unity  as $\mu$ goes to zero\footnote{This can be shown by differentiating $\frac{1-p_0(\mu)}{\mu}$ with respect to $\mu$ and then we know $\frac{d}{d\mu}\frac{1-p_0(\mu)}{\mu}<0\iff 1+\frac{4.5}{3.5}\mu<(1+\frac{\mu}{3.5})^{4.5}\label{eq:app}$, which holds since $1+\frac{4.5}{3.5}\mu$ is the tangent of $(1+\frac{\mu}{3.5})^{4.5}$ at zero and $(1+\frac{\mu}{3.5})^{4.5}$ is convex. Then by applying the L'H\^{o}pital rule on $\lim_{\mu\rightarrow 0}\frac{1-p_0(\mu)}{\mu}, i.e. \lim_{\mu\rightarrow 0}\frac{1-p_0(\mu)}{\mu}=-\lim_{\mu\rightarrow 0}p'_0(\mu)=1$. Thus $p_0(\mu)\rightarrow 1$ as $\mu\rightarrow 0$.}.  Fig. \ref{fig:user_rate} shows that the simulated and approximated results of  $\mathcal{R}_{\text{user},k}$, and it not only illustrates the accuracy of assuming the independence between  $\mathtt{SINR}$ and $N_b$ but also verifies that deploying more BSs per unit area with lower transmit power increases the average user rate due to the reduction in the cell load $\mu$. 
 
\section{Energy-Efficient Deployment Density of BSs}
The average cell and user rates have been specified in the previous section. Now we turn our attention on how much of the BS density is needed in order to maximize the energy efficiency of a BS.  The energy efficiency is defined by the ratio of the average cell rate to the average energy consumed by a BS. According to this energy efficiency definition, let $\eta_1$ and $\eta_2$ be the energy efficiency for the all-on energy scheme and the energy efficiency for the on-off energy scheme,  respectively. They can be written as follows
\begin{align}
\eta_1 &\defn \frac{\mathcal{R}_{\text{cell},1}}{P_{\text{on}}}
=\frac{[1-p_0(\mu)]\mathcal{C}_1}{P_0+\Delta P_t}, \label{eq:power_eff_1}\\
\eta_2 &\defn \frac{\mathcal{R}_{\text{cell},2}}{(1-p_0(\mu))P_{\text{on}}+p_0(\mu)P_{\text{off}}}\nonumber\\
&=\frac{[1-p_0(\mu)]\mathcal{C}_2}{(1-p_0(\mu))(P_0+\Delta P_t)+p_0(\mu)P_{\text{off}}}.\label{eq:power_eff_2}
\end{align}

Since our interest is in the optimal BS density that maximizes the energy efficiency, for a fixed user density the optimization problem of the BS density can be formulated as
\begin{equation}\label{Eqn:OptProbDensityBS}
  \text{maximize }\mathcal{\eta}_k\left( \lambda_b\right)\text{  subject to }\lambda_b\leq \lambda_u,
\end{equation}
for $k\in\{1,2\}$. The optimal solution of \eqref{Eqn:OptProbDensityBS} can be found as shown in the following lemma.
\begin{lemma}
 For a given user density $\lambda_u$, the optimization problem in \eqref{Eqn:OptProbDensityBS} can be transformed into a convex optimization problem of the cell load $\mu$. Therefore, the global maximizer $\lambda^*_{b,k}$ of \eqref{Eqn:OptProbDensityBS} uniquely exists.
\end{lemma}
\begin{IEEEproof}
See Appendix. 
\end{IEEEproof}
Although the optimal solution $\lambda^*_{b,k}$ exists, it is impossible to solve its closed form due to the high nonlinearity of $\eta_k(\lambda_b)$. In the following section, numerical results will verify that the energy efficiency $\eta_k$ is a convex function of the cell load $\mu$ and thus $\lambda^*_{b,k}$ can be found numerically.  

\section{Simulation Results}\label{Sec:NumericalResults}
This section presents some numerical results to evaluate the energy efficiency of the all-on and on-off power control schemes and verify that the optimal BS density for each power control scheme indeed exists. Suppose the small cell network consists of picocells so that the pico-to-user path loss model given in \cite{36.814} is adopted by incorporating average penetration loss $20\text{ dB}\times 0.8$ (due to 80\% users are indoor), antenna gain (5 dBi) \cite{36.872}. The BS power model is based on the picocell model in \cite{aue11}. The rest parameters of the small cell network for simulation are listed as follows: $C=4.33\times 10^{-6}$ m$^\alpha$, $\alpha=3.67$, $\delta=0.01$, $P_{r,\text{min}}=-100$ dBm (corresponding to the outage probability of $\mathcal{P}_1(5\text{ dB},-100\text{ dBm})=0.26$), $P_0 = 6.8$ watt, $\Delta = 4.0$, $P_\text{off}=4.3$ watt, noise power $\sigma_n^2=-95$ dBm (bandwidth of 10 MHz, noise power spectral density of -174 dBm and user noise figure of 9 dB). 

\begin{figure}[!t]
\centering
\includegraphics[width=3.6in,height=2.1in]{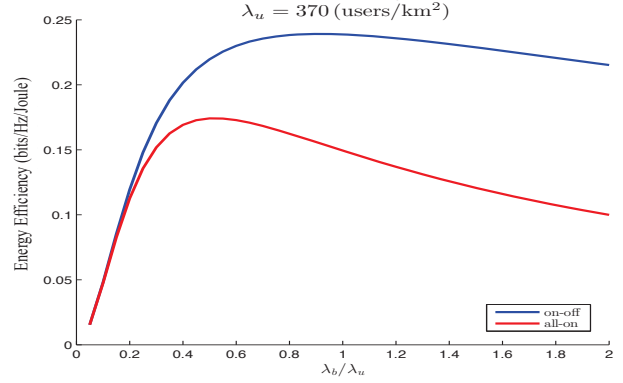}
\caption{Energy efficiency vs.  the BS density normalized by user density $\lambda_u=370\text{ users/km}^2$ for the all-on and on-off power control schemes}
\label{fig:power_efficiency}
\end{figure}

Fig. \ref{fig:power_efficiency} shows the results of energy efficiency versus the BS density for a given user density $\lambda_u=370 \text{ users/km}^2$ and verifies that there exists an optimal BS density achieving the maximum energy efficiency. By comparing the curves in the figure, on-off power control not only saves energy but also makes it possible to deploy more BSs per unit area to improve average user rate without compromising energy efficiency. Most importantly, its energy efficiency significantly outperforms that of all-on power control. Note that for the both power control schemes the optimal BS density that maximizes the energy efficiency indeed exists, for example, the highest energy efficiency for on-off power control is about 0.24 for $\lambda^*_{b,2}\approx 333\text{ BSs/km}^2$. However, it should be noted that the value of optimal BS density highly depends on the BS power model.

\begin{figure}[!t]
\centering
\includegraphics[width=3.6in,height=2.10in]{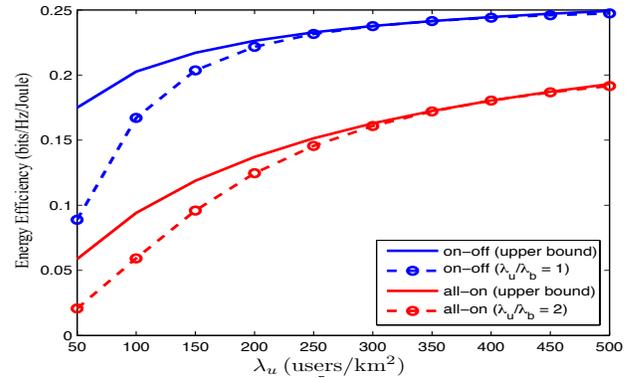}
\caption{Energy efficiency with the optimal BS density and energy efficiency with a non-optimal BS density under a fixed cell load}
\label{fig:power_efficiency_upper_bound}
\end{figure}

\begin{figure}[!t]
\centering
\includegraphics[width=3.6in,height=2.2in]{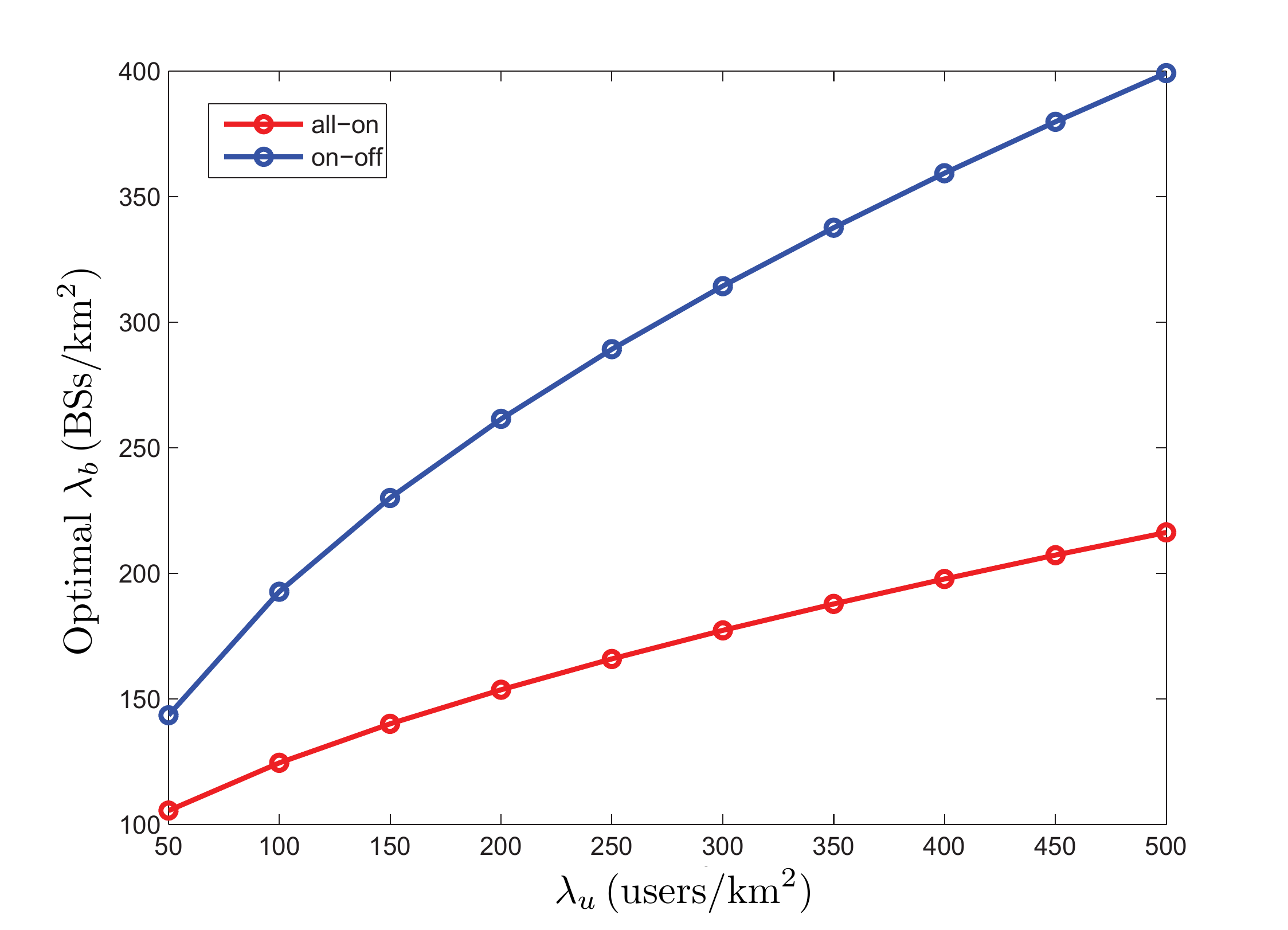}
\caption{Optimal BS density vs. user density}
\label{fig:optimal_den_b}
\end{figure}

 In Fig. \ref{fig:power_efficiency_upper_bound}, we illustrate the discrepancy between the maximum energy efficiency achieved by the optimal BS density and the energy efficiency achieved by the BS density under a fixed cell load. This discrepancy indicates that  for an arbitrary user density  the maximum energy efficiency cannot be achieved even though deploying BSs with a fixed cell load is kind of intuitive.  Fig. \ref{fig:optimal_den_b} plots the optimal BS density that achieves the maximum energy efficiency versus user density, and it essentially tells us how many BSs per unit area should be deployed for a given user density in order to attain the maximum energy efficiency.  Therefore, in order to improve the energy efficiency system operators can dynamically turn on or turn off BSs by referring to the statistics of the long-term density of active subscribers in the network. 
  
  %In Fig. \ref{fig:opt_criterion}, the optimality criterion clearly shows how optimal cell load $\mu_1^*$ varies with the system parameters. For example, to achieve maximum power efficiency in a densely populated area, a BS has to serve more users than a BS in a sparsely populated area.

%\begin{figure}[!t]
%\centering
%\includegraphics[width=1\linewidth]{./optimality_criterion}
%\caption{Optimality criterion (\ref{eq:opt_criterion}) for small cells with the all-on power scheme. The green line is for $\lambda_u=370$ km$^{-2}$.}
%\label{fig:opt_criterion}
%\end{figure}

\section{Conclusions}
In this paper, we investigate the problem of the optimal BS deployment for achieving high energy efficiency in a small cell network by proposing the all-on and on-off control schemes. The average user and cell rates affected by the void cell issue are explicitly calculated, respectively. Then the energy efficiency that is defined based on the average achievable cell rate is found for each of the control scheme, and it is shown as a convex function of the cell load so that the unique optimal BS density exists and can be numerically found. Simulation results show that on-off power control significantly outperforms the all-on power control in terms of energy efficiency if the BS density is optimally maintained for a known user density.   

\appendix
First consider the case of BSs with the all-on power control scheme. For $k=1$, (\ref{Eqn:OptProbDensityBS}) can be equivalently written as
\begin{equation}\label{Eqn:OptDenBSwOnOffPow}
\min_{1\geq\mu>0} \frac{P_0+\Delta  P_{r,0}\mu^{\frac{\alpha}{2}}/(\lambda^{\frac{\alpha}{2}}_uC)}{1-\left(1+3.5^{-1}\mu\right)^{-3.5}} 
\end{equation}
since $\mu=\frac{\lambda_u}{\lambda_b}$ and $\lambda_u$ is a given constant in this context. Hence, the optimization problem in  \eqref{Eqn:OptDenBSwOnOffPow} can be equivalently transformed to the following optimization problem:
 $$ \min_{1\geq \mu>0}\frac{g(\mu)}{q(\mu)},$$
 where $g(\mu)$ and $q(\mu)$ are
 \begin{align*}
 g(\mu) =P_0+\frac{\Delta  P_{r,0}\mu^{\frac{\alpha}{2}}}{\lambda^{\frac{\alpha}{2}}_uC}, q(\mu)=1-\frac{1}{\left(1+\mu/3.5\right)^{3.5}}.
 \end{align*}
Since $g(\mu)$ and $q(\mu)$ are both monotonically continuous and bounded functions for $\mu\in(0,1)$, there exists a minimum value of $1/\eta_1(\mu)$ according to the Bolzano–Weierstrass theorem. Namely,  the optimal BS density, $\lambda^*_{b,1}$, that minimizes $g(\mu)/q(\mu)$ exists.  
 
Now consider the other case of BSs with on-off power control. Since our objective is to show if there exists a solution in \eqref{Eqn:OptProbDensityBS}, we can instead use the following optimization problem for $k=2$:
 $$ \min_{\mu>0} \frac{v(\mu)}{w(\mu)}$$
 in which $v(\mu)$ and  $w(\mu)$ are 
 \begin{align*}
 v(\mu) &= \left(\lambda_u^{\frac{\alpha}{2}}P_0+\frac{\Delta P_{r,0}}{C}\mu^{\frac{\alpha}{2}}\right)-p_0(\mu)\\
 &\quad \, \left(\lambda_u^{\frac{\alpha}{2}}P_0+\frac{\Delta P_{r,0}}{C}\mu^{\frac{\alpha}{2}}-\lambda_u^{\frac{\alpha}{2}}P_{\text{off}}\right),\\
  w(\mu) &=[1-p_0(\mu)]e^{-\pi x\rho(2^t-1,\alpha)[1-p_0(\mu)]}.
 \end{align*}
 It is easy to show that $v(\mu)$ is monotonically decreasing for $\mu\in(0,1)$ since $\frac{\dif v}{\dif \mu}<0$. Since $h(x)e^{-ah(x)}$ is monotonically decreasing if $a>0$ and $h(x)>0$ is monotonically decreasing, $w(\mu)$ is monotonically decreasing because $1-p_0(\mu)$ is monotonically decreasing. Hence, $\min_{1\geq\mu>0} \frac{v(\mu)}{w(\mu)}$ has an optimal solution for $\mu$ and thus $\mu^*_2$ exists and thus $\lambda^*_{b,2}$ can be found readily.

\section*{Acknowledgment}
This work was supported by  MediaTek Inc. under research project 102C080.

\bibliographystyle{IEEEtran}
\bibliography{IEEEabrv,OptDepGreenPowCon}
\end{document}